\documentclass[11pt]{article}

\usepackage{color}

\usepackage{amsmath, amssymb,amsfonts,longtable,tabularx}
\usepackage{cite}
\usepackage{colortbl}
\usepackage[all]{xy}
\usepackage{epsfig}
\usepackage{subfigure}
\usepackage{graphics}
\usepackage{multirow}

\usepackage{setspace}
\usepackage{caption}
\usepackage{pifont}
\usepackage[table]{xcolor}

\usepackage{amsmath, amssymb,amsthm,bm,bbm,fullpage}
\usepackage{mathrsfs}
\usepackage{graphicx}

\newcommand{\ie}{\textit{i.e.}\,}
\newcommand{\eg}{\textit{e.g.}\,}
\newcommand{\nb}{\textit{n.b.}\,}
\newcommand{\defn}{:=}

\theoremstyle{remark}

	\usepackage[hmargin=2cm, vmargin=2cm]{geometry}

\date{}

\usepackage{cite}

\begin{document}

\begin{flushleft}
{\huge
\textbf{Evolutionary consequences of behavioral diversity}
}
\bigskip
\\
Alexander J. Stewart$^{1}$, Todd L. Parsons$^{2}$ and 
Joshua B. Plotkin$^{3}$
\\
\bigskip
\bigskip

$^1$ Department of Genetics, Environment and Evolution, University College London, London, UK
\\
$^2$ Laboratoire de Probabilit\'{e}s et Mod\`{e}les Al\'{e}atoires, CNRS UMR 7599, Universit\'{e} Pierre et Marie Curie, Paris 75005, France.
\\
$^3$ Department of Biology, University of Pennsylvania, Philadelphia, PA 19104, USA
\\
\end{flushleft}

\textbf{
Iterated games provide a framework to describe
social interactions among groups of individuals. Recent work stimulated by the
discovery of ``zero-determinant" strategies has rapidly expanded our ability to
analyze such interactions.
This body of work has primarily focused on games in which players face a simple
binary choice, to ``cooperate'' or ``defect''. Real individuals, however, often
exhibit behavioral diversity, varying their input to a social interaction both
qualitatively and quantitatively. Here we explore how access to a greater
diversity of behavioral choices impacts the evolution of social dynamics in finite
populations. We show that, in public goods games, some 
two-choice strategies can nonetheless resist invasion by all possible multi-choice invaders, even while engaging in relatively little punishment. 
We also show that access to greater behavioral choice results in more ``rugged '' fitness landscapes, with populations able to stabilize cooperation
at multiple levels of investment, such that choice facilitates cooperation when returns on investments are low, but hinders cooperation when returns on investments are high. Finally, we analyze iterated rock-paper-scissors games, whose non-transitive payoff structure means unilateral control is difficult and zero-determinant strategies do not exist in general. Despite this, we find that a large portion of multi-choice strategies can invade and resist invasion by strategies that lack behavioral diversity -- so that even well-mixed populations will tend to evolve behavioral diversity.}
\\
\\
Diversity in social behaviors, not only in humans but across all domains
of life, presents a daunting challenge to researchers who work to explain and
predict individual social interactions or their evolution in populations.
Iterated games provide a framework to approach this task, but determining the
outcome of such games under even moderately complex, realistic assumptions -- such
as memory of past interactions
\cite{Hauert,Milinski:1998aa,Stewart:2016aa,DBLP:journals/tec/LiK14,Suzuki:2004aa,Suzuki:2013aa,Rand1},
signaling of intentions, indirect reciprocity or identity
\cite{Rand3,Rand:2012aa,Nowak:2006ys,Jordan:2016aa,Hauert:2007fk,NowakIndirect,Chen:2012aa,Bergstrom:2002aa},
or a heterogeneous network of interactions
\cite{Rand2,Rand:2013aa,Nishi:2015aa,Lieberman:2005aa,Ohtsuki:2006aa,Ohtsuki:2007aa,Ohtsuki:2007ab,Ohtsuki:2006ab,Hauert:2004aa}
-- is exceedingly difficult.

The discovery of zero-determinant (ZD) strategies \cite{Press:2012fk} has
stimulated rapid advances in our ability to analyse iterated games
\cite{Akin,Akin2,Hilbe:2014aa,Stewart:2012ys,Stewart:2014aa,Stewart:2016aa,McAvoy,Hilbe:2015ab,Pan:2015aa},
leading to new understanding of how one individual can influence the longterm
outcome of a pairwise social interaction, the evolutionary potential for cooperation, the prospects for
generosity and extortion among groups, and the role of memory in social dynamics
\cite{Hilbe:2013aa,Hilbe:2013uq,Stewart:2013fk,StewartGames,Hilbe:2015aa}. These
advances all rest on a key mathematical insight: the outcome of iterated games can
be easily understood when players' strategies, even those of startling complexity
\cite{Stewart:2016aa,Hilbe:2014aa,Pan:2015aa}, are viewed in the right coordinate
system. This coordinate system was suggested by the discovery of ZD strategies and
developed fully by Akin \cite{Akin} and others
\cite{Hilbe:2014aa,Hilbe:2015ab,McAvoy,Stewart:2014aa,Stewart:2016aa}.  ZD
strategies have also been generalized to two-player games with arbitrary actions
spaces \cite{McAvoy}.  Here, we study evolutionary dynamics in the full space of
memory-1 strategies in a population of players with access to multiple behavioral
choices, including games for which no ZD strategies exist at all.

Many game-theoretic studies of social behavior, although by no means all 
\cite{McAvoy,Killingback:2002aa,Doebeli:2004aa}, constrain players to a binary
behavioral choice such as ``cooperate'' or ``defect''
\cite{Nowak:2006ly,Sigmund:2010ve}.  Other studies, particularly those looking at
social evolution, constrain players to a single type of behavioral strategy, but
allow for a continuum of behavioral choices -- e.g. the option to contribute an
arbitrary amount of effort to an obligately cooperative interaction
\cite{Killingback:2002aa,Doebeli:2004aa}. In general, and especially in the case
of human interactions, individuals have access to both a wide variety of
behavioral choices, and to a complex decision making process among these choices.
Here we bridge this gap and study how the diversity of behavioral choices impacts
the evolution of decision making in a replicating population, focusing on the
prospects for cooperation and for the maintenance of behavioral diversity.

We develop a framework for analyzing iterated games in which players have an
arbitrary number of behavioral choices and an arbitrary memory-1 strategy for
choosing among them.  We apply this framework to study the effect of a large
behavioral repertoire on the evolution of cooperation in public goods games.  We
show that increasing the number of investment levels available to a player can
either facilitate or hinder the evolution of cooperation in a population,
depending on the ratio of individual costs to public benefits in the game.  We
apply the same framework to study games with non-transitive payoff structures,
such as rock-paper-scissors, and we show that, while ZD strategies in general do
not exist for such games, nonetheless memory-1 strategies exist that ensure the
maintenance of behavioral diversity, in which players make use of all the choices
available to them.

\section*{Methods and Results}

Players in an iterated game repeatedly choose from a fixed set of possible
actions.  Depending on the choice she makes, and the choices her opponents make, a
player receives a certain payoff each round.  The process by which a player
determines her choice each round is called her strategy. A strategy may in general
take into account a wide variety of information about the environment, memory of
prior interactions between players, an opponent's identity, his social signals etc
\cite{Hauert,Milinski:1998aa,Stewart:2016aa,DBLP:journals/tec/LiK14,Suzuki:2004aa,Suzuki:2013aa,Nowak:2006ys,Hauert:2007fk,NowakIndirect,Chen:2012aa,Lieberman:2005aa,Ohtsuki:2006aa,Ohtsuki:2007aa,Ohtsuki:2007ab,Ohtsuki:2006ab,Hauert:2004aa,Bergstrom:2002aa}.
Here we restrict our analysis to two-player, simultaneous infinitely iterated games in which
a player chooses from among $d$ possible actions using a memory-1 strategy, which
takes account only the immediately preceding interaction between her and her
opponent. Although memory-1 strategies may seem restrictive, in fact a strategy
that is a Nash equilibrium or evolutionary robust against all memory-1 strategies is
also robust against all longer memory strategies as well (see SI and
\cite{Press:2012fk,Pan:2015aa,Stewart:2014aa,Stewart:2016aa}).

A memory-1 strategy is specified by choosing $d^2$ probabilities for each possible
action $i$, denoted $p^i_{jk}$, which specify the chance the player executes that
action in a round of play, given that she made choice $j$ and her opponent made
choice $k$ in the preceding round.  Each probability can be chosen independently,
save for the constraint that the sum across actions $\sum_{i=1}^dp^i_{jk}=1$ must
hold.  We study the evolution of social behavior by analyzing the composition of
such strategies in a replicating population over time.  In an evolving population
the reproductive success of a player depends on the total payoff she receives in
pairwise interactions with other members of the population \cite{Traulsen:2006zr}.
We study how strategy evolution is affected by the number and by the types of
behavioral choices available to individuals.

We study two qualitatively different behavioral choices that players can make:
different \textit{sizes} of contributions and different \textit{types} of
contributions to social interactions (Figure 1). If players can vary the size of
the contribution they make to a social interaction, this means that they alter the
degree of their participation but not the qualitative nature of the interaction.
For example, in a public goods game, a player may choose to contribute an amount
$C$ to the public good, or $2C$, or $3C$ etc. In contrast, when players can vary
the type of contribution they make, this can change the qualitative nature of the
social interaction. For example, in a game of rock-paper-scissors the different
behavioral choices result in qualitatively different social interactions -- rock
beats scissors, but scissors beats paper, etc. Such qualitative differences can
lead to non-transitive payoffs and correspondingly complex social and evolutionary
dynamics
\cite{Kerr:2002kx,Lively,Reichenbach:2007aa,Reichenbach:2007ab,Szolnoki:2014aa,Szczesny:2014aa,Bergstrom:2015aa}.

Here we study both kinds of behavioral choice, differences in size and type, and
their effects on the evolution of strategies in a population. We analyze
well-mixed, finite populations of $N$ players reproducing according to a copying
process, in which a player $X$ copies her opponent $Y$'s strategy with probability
$1/(1+\exp\left[\sigma(S_{x}-S_{y})\right])$ where $\sigma$ scales the strength of
selection and $S_{x}$ is the average payoff received by player $X$ from her social
interactions with each of the $N-1$ other members of the population
\cite{Nowak:2006ly,Traulsen:2006zr}, which corresponds to the fitness associated
with the strategy given the current composition of the population. For a single
invader $Y$ in a population otherwise composed of strategy $X$, this means
$S_{y}=S_{yx}$ and $S_x=\frac{N-2}{N-1}S_{xx}+\frac{1}{N-1}S_{xy}$

\begin{figure}[h!] \centering \includegraphics[scale=0.3]{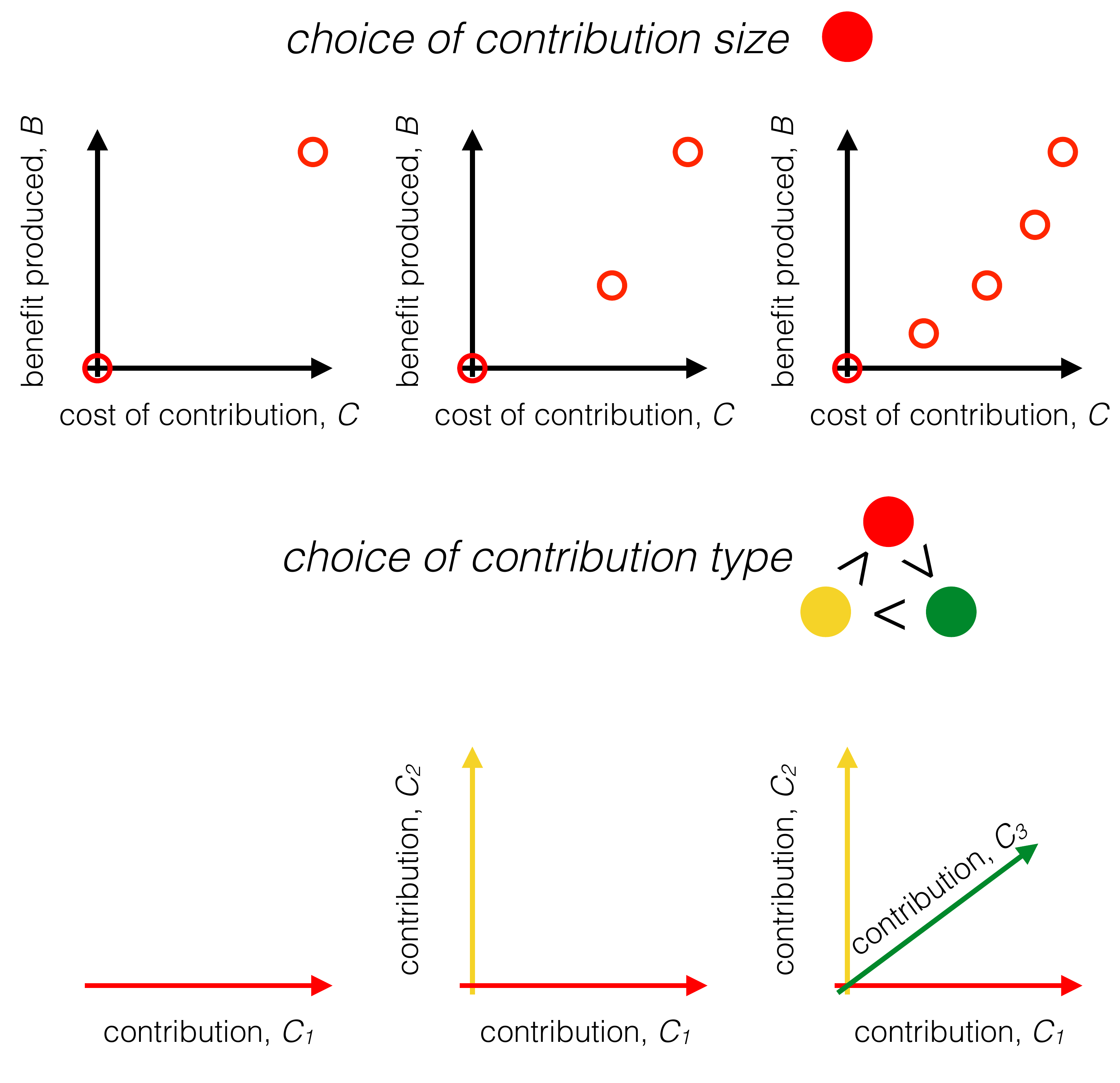}
\caption{\small  Two ways to expand the behavioral repertoire in iterated
games.  (Top) In a public goods game a player contributes to a public pool at some
cost to herself, and she receives a benefit based on the contributions of all
players in the game.  In a simple two-choice game, such as the Prisoner's Dilemma,
players face a binary choice, to cooperate and contribute cost $C$ or to defect
and contribute nothing.  At the other extreme, in a continuous game, players have
an unlimited number of options and may contribute any amount.  What happens to the
evolution of social behavior as the numbers of choices increases?  Is it
beneficial for a population to have access to more choices in a public goods game?
(Bottom) Players may also choose between qualitatively different
\textit{types} of contributions to social interactions. For example,
unicellular organisms may produce pathogens, social signals, public goods or all
three \cite{Kerr:2002kx,Allen:2013aa,Cordero:2012vn,Kelsic:2015aa}.  Qualitatively
different behavioral options produce complex payoff structures, such as the
non-transitive rock-paper-scissors interactions
\cite{Kerr:2002kx,Lively,Reichenbach:2007aa,Reichenbach:2007ab}. What happens to
the evolution of social behavior as the types of contributions to social
interactions expand?  Is it better to maintain a diversity of behavioral options,
or to restrict to a single type of contribution?} \end{figure}

\subsection*{The outcome of an infinitely iterated $d$-choice game:}

To analyse social evolution in multi-choice iterated games we must first calculate
the expected longterm payoff $S_{xy}$ of an arbitrary player $X$ facing an
arbitrary opponent $Y$.  To do this, we will generalize an approach used for
two-choice two-player games, in which a player's memory-1 strategy $\mathbf{p}$ is
represented in an alternate coordinate system \cite{Akin} so that the outcome of
the repeated game can be determined with relative ease.  For a $d$-choice
two-player game, the probability that a focal player chooses action $i$, given
that she played action $j$ and her opponent action $k$ in the preceding round, is
denoted $p^i_{jk}$.  For each action $1\leq i < d$ there are $d^2$ independent
probabilities, corresponding to each possible outcome of the preceding round.  In
the alternate coordinate system we construct (see SI), the probabilities
$p^i_{jk}$ are written as linear combinations of the payoff $R_{jk}$ the focal
player received in the preceding round, times a coefficient $\chi^i$; the payoff
$R_{kj}$ her opponent received, times a coefficient $\phi^i$; the number of times
she played action $i$ within her memory (which is one or zero for a memory-1
strategy); a baseline rate of playing action $i$, denoted $\kappa^i$; and $d^2-3$
additional terms that depend on the specific outcome of the preceding round,
denoted $\lambda^i_{jk}$.  This choice of coordinate system enforces the following
relationship between the longterm average payoffs received by the two players: 
\begin{equation} \phi^i
S_{yx}-\chi^i S_{xy}-(\phi^i-\chi^i)\kappa^i
+\sum_{j=1}^d\sum_{k=1}^d\lambda^i_{jk}v_{jk}=0 \end{equation} 
\\ where $v_{jk}$ denotes the equilibrium rate of action $jk$, and where we fix
the values of three of the $\lambda^i_{jk}$ to ensure a system of $d^2$
coordinates (see SI).  Note there are $d-1$ such equations, one for each
behavioral choice $1\leq i <d $. A ZD strategy of the type studied in
\cite{McAvoy} can be recovered by setting all $\lambda^i_{jk}=0$.  However the
constraint that $p^i_{jk}\in[0,1]$ implies that the ZD condition does not always
produce a viable strategy, as in the case of a rock-paper-scissors game discussed
below.


\subsection*{Choosing how much to contribute to a public good:}

We will use the relationship between two players scores (Eq.~1) to analyse the
evolution and stability of cooperative behaviors in multi-choice public goods
games, played in a finite population. In the two-player public goods game each
player chooses an investment level, $C$, which produces a corresponding amount of
public benefit that is then shared equally between both players, regardless of
their investment choices.  In general, if a player invests $C_j$ and her opponent
$C_k$ the public benefit produced is determined by a function $B(C_j+C_k)$, so
that her net payoff is $B(C_j+C_k)/2-C_j$ while her opponent's payoff is
$B(C_j+C_k)/2-C_k$.  Two-choice public goods games have been studied extensively,
producing a clear understanding of the cooperative equilibria that exist in
populations
\cite{Stewart:2014aa,Stewart:2013fk,Stewart:2016aa,Hilbe:2013uq,Hilbe:2013aa,Akin,Akin2}.

A wide variety of evolutionary robust memory-1 strategies exist for two-choice
public goods games. The character and evolvability of these strategies have been
explored in detail
\cite{Stewart:2014aa,Stewart:2013fk,Stewart:2016aa,Hilbe:2013aa,Imhof:2007uq,Nowak:1993vn,Axelrod:1981kx}.
But the assumption of only two investment levels -- of two behavioral choices --
is unrealistic for many applications. Even if a player adopts such a two-choice
strategy, there is in general no reason for her opponent to do the same. Thus we
begin our analysis by asking whether a cooperative, two-choice, memory-1 strategy
resident in a population can resist invasion against players who can make
arbitrary investment choices. 

For simplicity, we will focus here on a linear relationship between costs and
benefits of investment in the public good, so that $B=rC$ where values $1<r<2$
produce a social dilemma in which mutual cooperation is beneficial but each player
has an incentive to defect.  The more general case, with non-linear functional
relationships, is described in the Supporting Information.

For linear benefits, a two-choice strategy is completely defined by 

\begin{eqnarray*}
p_{1i}&=&1-\left((\phi
-\chi )(r(C_1+C_i)/2-\kappa)-\phi C_i+\chi C_1+\lambda_{1i}\right)\\
p_{2i}&=&-\left((\phi -\chi)(r(C_2+C_i)/2-\kappa)-\phi C_i+\chi
C_2+\lambda_{2i}\right)
\end{eqnarray*}

where the index $i$ corresponds to an opponent who
invests $C_i$, which in general can take any
non-negative value. Here we choose the
boundary conditions $\lambda_{11}=\lambda_{22}=0$ and $\lambda_{12}=\lambda_{21}$,
and from Eq.~1 we obtain the following relationship between two players'
longterm payoffs
\begin{equation*}
\phi S_{yx}-\chi S_{xy}-(\phi-\chi)\kappa+\lambda_{12}(v_{12}+v_{21})+\sum_{j=3}^d(\lambda_{1j}v_{1j}+\lambda_{2j}v_{2j})=0
\end{equation*}
\\
When player $Y$ is constrained to the same two choices as player $X$, 
then this relationship reduces precisely 
to the relationship for two-player, two-choice games discussed in \cite{Akin,Press:2012fk,Stewart:2013fk,Stewart:2014aa}. However, we will consider the more general case when player
$Y$ has access to different, and possibly more, investment choices than $X$.
In general, a strategy $X$ resident in a population of $N$
players can resist selective invasion by a mutant $Y$ iff
\begin{equation*}
S_{yx}<\frac{N-2}{N-1}S_{xx}+\frac{1}{N-1}S_{xy}
\end{equation*}
\\
where $S_{xx}$ is the longterm payoff of the resident strategy against itself.
A cooperative two-choice strategy by definition has
$S_{xx}=(r-1)C_{2}$, i.e.~it stabilizes cooperative behavior at equilibrium, 
with both players choosing to invest the maximum public good they can contribute.

Using the relationships above we can derive the following conditions for a
two-choice cooperative strategy to be \textit{universally robust} to invasion --
that is, robust against all invaders $Y$, who can make an arbitrary number of
different investment choices, including values above $C_2$ or below $C_1$
(see SI):
\small
\begin{align}
\nonumber  &\mathcal{C}^{d}_{s}=\Bigg\{(p_{11},p_{12},\ldots,p_{1d},p_{21},p_{22},\ldots,p_{2d})\bigg|p_{11}=1,\\
\nonumber &p_{1j}<1-\frac{N-2}{N}(1-p_{12}+p_{21})\frac{1-c^*}{1-c}\left[\frac{N-1}{N-2}-\frac{r}{2}\right], \\
\nonumber  &p_{2j}<
\frac{N-2}{N}(1-p_{12}+p_{21})\left[\frac{r}{2}-\left(\frac{N-1}{N-2}-\frac{r}{2}\right)\frac{1-c^*}{1-c}+\frac{1}{N-2}\right],\\
\nonumber &p_{1j}<1-\frac{p_{22}}{r-1}\frac{1-c^*}{1-c}\left[\frac{N-1}{N-2}-\frac{r}{2}\right]\\
\nonumber &p_{2j}<\frac{p_{22}}{r-1}\left[\frac{r}{2}-\left(\frac{N-1}{N-2}-\frac{r}{2}\right)\frac{1-c^*}{1-c}+\frac{1}{N-2}\right]
\Bigg\},
\end{align}
\normalsize
\begin{equation}
\end{equation}
\\
where we have set $c=C_1/C_2$ and $c^*=C_j/C_2$. All four of these inequalities
are hardest to satisfy when $c^*=0$, i.e. when an invader does not invest at all in
the public good (although this is not necessarily the case when benefits vary
non-linearly with costs -- see Supporting Information). Using this fact, alongside
the requirement that a strategy be viable (i.e. $p_{ij}\in[0,1], \ \ \forall
i,j$), we can derive the following necessary and sufficient 
condition for the existence of a two-choice
strategy that is universally robust:

\begin{equation}
\frac{C_{1}}{C_{2}}<\frac{r-1}{\frac{r}{2}+\frac{1}{N-2}}
\end{equation}
\\
If (and only if) Eq. 3 is satisfied, then there exists a two-choice strategy that enforces
cooperation at some level when resident in a population, and that 
resists invasion by any invader, regardless of the invader's ability to choose different investment levels.

Eq.~3 offers insight into the degree of punishment that a resident cooperative
strategy must be prepared to wield, in order to remain robust against all invaders
(Fig.~ 2).  A resident strategy can punish a non-cooperative invader by
reducing her investment in the public good from $C_2$ to $C_1$. If $C_1$ is only
slightly smaller than $C_2$ then the resident strategy has a limited capacity to
punish invaders. Wheres if $C_1$ is much less than $C_2$ the resident strategy has
a greater capacity for punishment.  The critical question is how much capacity for
punishment, quantified by the ratio of $C_1$ and $C_2$, is required to ensure that
the resident cooperator can be robust against all invaders (who can make arbitrary
investments, outside of those available to the resident). The answer to this question
is shown in Figure 2, which quantifies the minimum reduction in public investment
that a cooperative two-choice strategy must make in order to be universally
robust.
As might be expected from Eq. 3, larger ratios of public benefit to individual cost $r$ and larger
population sizes $N$ mean that smaller reductions in public investment are
sufficient for universal robustness of the resident cooperator. And as Fig. 2
shows, for a wide range of parameters a population can enjoy robust cooperation
using a simple two-choice strategy with only moderate threat of punishment, e.g.
$C_1$ no less than than one-half of $C_2$.

\begin{figure}[h!] \centering
\includegraphics[scale=0.25]{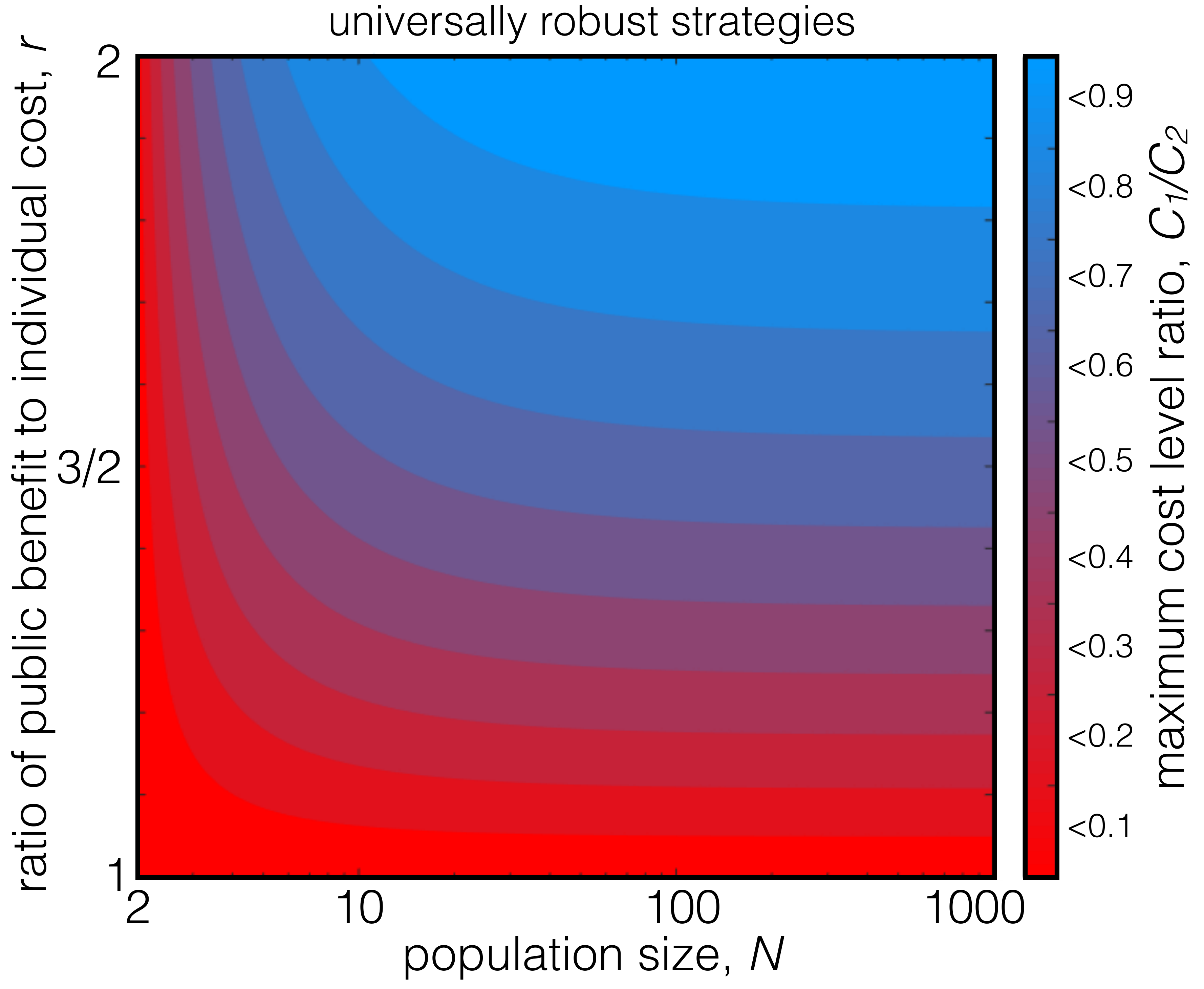}
\caption{\small  When are simple two-choice strategies robust against all
multi-choice invaders in public goods games? We considered the evolutionary
robustness of two-choice strategies, in which players iteratively choose to invest
amount $C_1$ or $C_2>C_1$ to produce a public benefit $B$ proportional to the
total investment of both players, $B=rC$.  Cooperative strategies limited to two
investment choices can be evolutionary robust against all invaders, who may invest
an arbitrary amount $C\neq C_1,C_2$, provided the strategy has sufficient
opportunity to punish a defector -- that is, provided $C_1$ is sufficiently
smaller than $C_2$. We determined (Eq.~4) the largest ratio of investment levels,
$C_2/C_1$, that permits universally robust cooperative two-choice strategies, as a
function of the the population size, $N$, and the public return on individual
investment, $r$.  Colors are gradated in 10\% intervals, so that the light blue
region indicates a two-choice player can choose a strategy that maintains robust
cooperation while engaging in relatively little punishment, by reducing her
investment to only 90\% of its maximum. The bright red region indicates that a
two-choice player must have access to a high degree of punishment, $C_1$ much less
than $C_2$, in order to maintain cooperation.  As described in Eq. 4, the figure
can alternatively be interpreted as the proportion of pairs of investment levels
used by a $d$-choice player that produce a robust sub-optimal fitness peak, and
thus represents a lower bound on the ``ruggedness'' of the fitness landscape
experienced by a population of $d$-choice players.} \end{figure}

\subsection*{Evolutionary consequences of multiple investment choices:}

We now turn our attention to the implications of these results for an evolving
population of players who can make $d>2$ choices for investment in the public
good.  We assume a discrete series of $d+1$ investment levels, from 0 to the
maximum $C_{max}$, so that subsequent levels of investment differ by $C_{max}/d$.
When $d$ is large, players have more options for investment, between the
fixed minimum value zero and fixed maximum value $C_{max}$.

Because all two-choice strategies form a subset of $d$-choice strategies, an
evolving population of $d$-choice players has access to, at minimum, all
evolutionary robust two-choice strategies. Thus, unlike in the two-choice case,
where there are only three qualitatively distinct types of evolutionary robust
strategies \cite{Stewart:2014aa}, a $d$-choice population may result in many different classes of
evolutionary robust outcomes, most of which are sub-optimal in the sense that they 
produce less public good than the global maximum, $rC_{max}$.

We can place a lower bound on how many such sub-optimal, but 
evolutionary robust, outcomes are
possible when players have $d+1$ choices. 
Any given pair of investment levels $C_i$ and $C_j$, with
$i>j$, can be a robust two-choice strategy provided $C_i$ and $C_j$ satisfy Eq. 4. 
Thus all pairs of
investment levels $j<\frac{r-1}{\frac{r}{2}+\frac{1}{N-2}}i$ have viable robust
two-choice strategies associated with them; and for a $d+1$-choice game the total number of
such evolutionary robust but sub-optimal strategies, $P_r$, satisfies
\begin{equation}
P_r>\left(\frac{r-1}{\frac{r}{2}+\frac{1}{N-2}}\right)\frac{d(d+1)}{2}.
\end{equation}
\\
Thus the number of sub-optimal evolutionary robust outcomes grows at least quadratically with
the number of investment levels available to individuals.

Fig. 2 can now be re-interpreted as showing the proportion of pairs of investment
levels that can produce a robust, sub-optimal two-choice strategy for a population
of $d+1$-choice players.  To put these results in perspective, if players are
allowed $d=100$ investment choices, with return on investment $r=3/2$, then in a
population size $N=1,000$ there are at least $3.6\times10^3$ robust strategies
that fail to maximize the total public good -- resulting in an extremely
``rugged'' fitness landscape and a large number of sub-optimal evolutionary
outcomes. By contrast, with only $d=2$ choices, there are at most two sub-optimal
evolutionary robust outcomes \cite{Stewart:2014aa}. 

We have seen that increasing the number of available choices to players, between a
fixed minimum and maximum investment level, has the potential to produce
sub-optimal but evolutionary robust outcomes. To test how the number of available
choices impacts evolutionary dynamics in a population, we ran evolutionary
simulations under weak mutation \cite{Stewart:2013fk}, with mutants drawn uniformly from all
$d$-choice memory-1 strategies.  We compared the mean payoffs received by
populations constrained to $d=2$ choices, to the mean payoffs in populations with
access to $d=11$ choices (Figure 3). The results are striking: when ratios of public benefit to individual cost are low, so that robust strategies are rare (Eqs.~2-3), the population that
has $d=11$ investment choices evolves a higher mean payoff than the $d=2$ choice
population -- because a greater number of robust cooperative strategies provides 
an advantage. But when ratios of public benefit to individual cost are higher, so that robust
strategies are more common, the 11-choice population evolves a lower mean payoff
than the 2-choice population -- because the huge number of sub-optimal robust
strategies causes the 11-choice population to ``get stuck'' and fail to maximize
its evolutionary potential. Thus, increasing the number of investment options,
between a fixed minimum and maximum, can either facilitate or hinder cooperative
interactions in a population.

\begin{figure}[h!] \centering \includegraphics[scale=0.25]{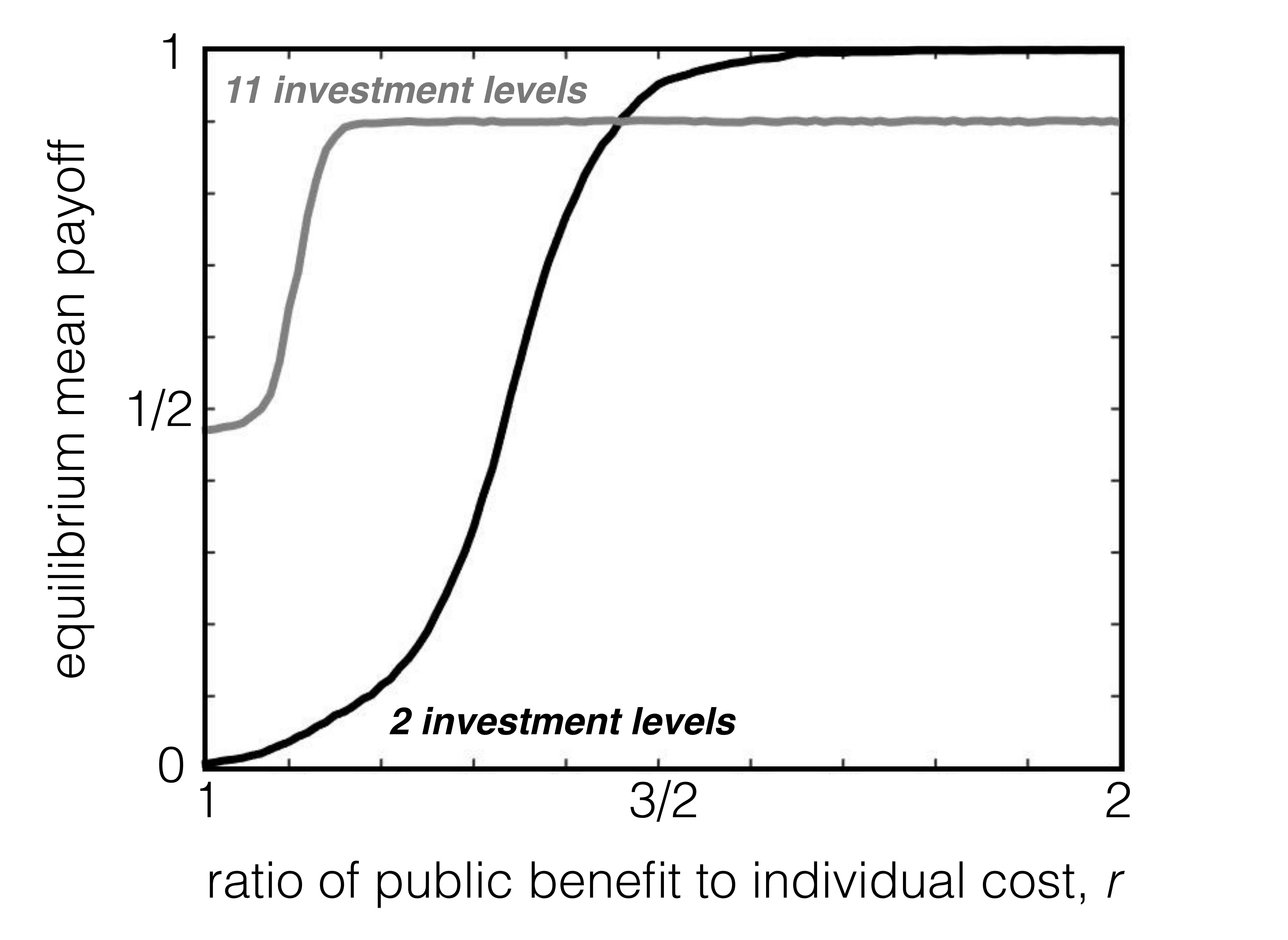}
\caption{\small Does a larger behavioral repertoire make cooperation easier to
evolve?  We evolved a well-mixed population of $N=100$ haploid, asexual
individuals reproducing according to the copying process \cite{Traulsen:2006zr} with an
individual's fitness determined by playing pairwise iterated public goods games.
We calculated ensemble mean fitness across $10^5$ replicate populations, each
evolved under weak mutation for at least $10^6$ fixation events.  We compared
populations with only two investment choices available, $C_1=0$ and $C_2=1$,
versus populations in which players could choose among 11 levels of investment,
between $0$ and $1$ in increments of $0.1$. In both cases evolution occurred on
the full set of memory-1 strategies. When the ratio of public benefit to individual cost is small,
two-choice populations evolve to low mean fitness and exhibit little cooperation,
whereas 11-choice populations evolve higher fitness and higher levels of investment in
the public good.  However, when the ratio of public benefit to individual cost is higher two-choice
populations evolve strategies that maximize the public good, whereas 11-choice
populations are less cooperative and receive roughly 10\% payoff reduction compared
to the two-choice case. Thus, a larger repertoire of behavioral options can either
facilitate or impede the evolution of cooperation, depending upon the public return
on individual investment.} \end{figure}

\subsection*{Non-transitive payoff structures:}

So far we have focused on multiple options for investment and its impact on the
evolution of cooperative behaviors in public goods games.  But the co-ordinate
system we have introduced for studying multi-choice iterated games, and the
resulting relationship between two players' scores (Eq.~1), applies generally, and
so it can be applied to study many other questions in evolutionary game theory.
Among the most interesting questions occur with only $d=3$ choices, but with
non-transitive payoffs, where the evolutionary dynamics are complex and the impact
of repeated interactions remains unclear \cite{Kerr:2002kx,Lively,Reichenbach:2007aa,Reichenbach:2007ab,Szolnoki:2014aa,Szczesny:2014aa,Bergstrom:2015aa}.


Games with non-transitive payoff structures, such as rock-paper-scissors,
describe social dynamics without any strict hierarchy of behaviors. Individuals
can invest in qualitatively different types of behavior, which dominate in some
social interactions but lose out in others.  Such non-transitive interactions have
been observed in a range of biological systems, from communities of
\textit{Escherichia coli} species \cite{Kerr:2002kx}, to mating competition among male
side-blotched lizards \textit{Uta stansburiana} \cite{Lively}.  Rock-paper-scissors
interactions are well known in ecology as having important consequences for the
maintenance of biodiversity: in well mixed populations playing the one-shot game,
diversity is often lost, whereas in spatially distributed populations multiple
strategies can be stably maintained  \cite{Reichenbach:2007aa,Reichenbach:2007ab}.  Here we analyse the equivalent
problem for the maintenance of diversity in evolving populations of players
who engage in iterated non-transitive interactions. 

We will assess the potential for maintaining \textit{behavioral diversity} in a
population playing an iterated rock-paper-scissors game -- that is, we look for
strategies that can resist invasion by players who employ a single behavioral
choice (1=rock, 2=paper or 3=scissors). We assume that, in any given interaction,
a fixed benefit $B$ is at stake, and players invest a cost $C_1$, $C_2$ or $C_3$
to execute the corresponding behavioral choice. Under the rock-paper-scissors game
we then have payoffs $R_{13}=B-C_1$, $R_{21}=B-C_2$, $R_{32}=B-C_3$,
$R_{31}=-C_3$, $R_{12}=-C_1$ and $R_{23}=-C_2$. When two players make the same
choice we assume they receive equal payoff: $R_{11}=B/2-C_1$, $R_{22}=B/2-C_1$ and
$R_{33}=B/2-C_1$.

We first consider the case of a completely symmetric game of
rock-paper-scissors, with $C_1=C_2=C_3=C$. In this case a given
round of the game has only three distinct outcomes for a player: win (+), lose (-)
or draw (o). A player's memory-1 strategy can be thought of as the
probability that she plays, for example, a move that would have won in the
preceding round, given that she lost. We write this probability $p^+_-$. Similarly
$p^-_-$ is the probability she plays the same move that lost the preceding round;
and $p^o_-$ is the probability that she plays the move that would have resulted in
a draw. This symmetric strategy is thus composed of 9 probabilities, which are written 
in our alternative coordinate system as:

\begin{eqnarray*}
p^o_{o}&=&1-(\phi-\chi)\left(B/2-C-\kappa\right)\\
p^-_{-}&=&1-\left(\phi(B-C)+\chi C-(\phi-\chi)\kappa\right)\\
p^+_{+}&=&1+\left(\phi C+\chi(B-C)+(\phi-\chi)\kappa\right)\\
p^o_{+}&=&\lambda^o_{+}+\left(\phi C+\chi(B-C)+(\phi-\chi)\kappa\right)\\
p^-_{o}&=&\lambda^-_{o}-(\phi-\chi)\left(B/2-C-\kappa\right)\\
p^+_{-}&=&\lambda^+_{-}-\left(\phi(B-C)+\chi C-(\phi-\chi)\kappa\right)\\
p^o_{-}&=&\lambda^o_{-}-\left(\phi(B-C)+\chi C-(\phi-\chi)\kappa\right)\\
p^-_{+}&=&\lambda^-_{+}+\left(\phi C+\chi(B-C)+(\phi-\chi)\kappa\right)\\
p^+_{o}&=&\lambda^+_{o}-(\phi-\chi)\left(B/2-C-\kappa\right)\\
\end{eqnarray*}
\\
where we have set $\lambda^o_{o}=\lambda^+_{+}=\lambda^-_{-}=0$ as a boundary
condition. We see immediately from this that there exists no viable ZD strategy, for which
$\lambda^i_j=0, \ \ \forall i,j$, unless we
also set $\kappa=\chi=\phi=0$ to produce the singular ``repeat'' strategy \cite{Press:2012fk}.
Nonetheless, we can still analyse the outcome of iterated rock-paper-scissors
games using this coordinate system.

\subsection*{Maintaining behavioral diversity in a game of rock-paper-scissors:}

The symmetric, iterated rock-paper-scissors game is simple to analyse, because
payoff is conserved, meaning that the sum of two interacting players' payoffs is
constant, $S_{xy}+S_{yx}=B-2C$. Thus the expected fitness of a population is
independent of the strategy that is resident, and $S_{xx}=B/2-C$ holds for all
strategies $X$. It might seem unlikely, then, that behavioral diversity offers any
advantage in this situation. After all,  
a player who uses a strategy that
employs only rock, paper or scissors produces no higher mean fitness at the population level
than a player who always uses rock. To determine whether this intuition is
correct, and non-transitive payoffs lead inevitably to a loss of behavioral diversity, we evaluated the
conditions for a strategy to resist selective invasion by a player who always
uses the same move. Such strategies do
indeed exist, and satisfy the following inequality:

\begin{equation}
p^-_{o} (1-p^-_{-} - p^-_{+} ) >  p^+_{o}(1-p^+_{+} - p^+_{-}).
\end{equation} 
\\
As one might hope, strategies that tend to switch to the move that would have won
in the preceding round -- corresponding to larger values of $p^+_{o}$, $p^+_-$,
$p^+_+$  and smaller values of $p^-_{o}$, $p^-_-$, $p^-_+$  -- tend to be
evolutionary robust.
However Eq. 6 also provides a more valuable insight, if we calculate the overall
robustness of memory-1 strategies to the loss of behavioral
diversity. To do this we calculate the probability that a randomly drawn
memory-1 strategy satisfies Eq. 6, which reveals that fully 50\% of such strategies
maintain behavioral diversity in the completely symmetric rock-paper-scissors
game (Figure 4). Furthermore, due to symmetry, the condition for a
new strategy to invade a resident is simply $S_{yx}>S_{xy}$ (see
SI). And so if a resident can resist invasion against a
particular invader, it can also invade a population in which that invader is
resident. Thus 50\% of strategies can successfully invade in a
population that lacks behavioral diversity -- so that behavioral diversity is both highly
evolvable and easy to maintain in the iterated rock-paper-scissors game, even in a
well-mixed population -- in sharp contrast to the one-shot game.

We can also assess the robustness of behavioral diversity when the symmetry of the
game is broken, so that $C_1\neq C_2\neq C_3$. In Figure 4a we numerically
calculate the overall robustness of randomly drawn strategies as a function of the
costs $C_1/C_3$ and $C_2/C_3$ keeping $B$ and $C_3$ fixed. We find that, for a
wide range of costs, including in some cases with $B<C$, behavioral diversity can
be maintained with relative ease in an evolving population (Fig.~4).  

\begin{figure*}[h!] \centering \includegraphics[scale=0.22]{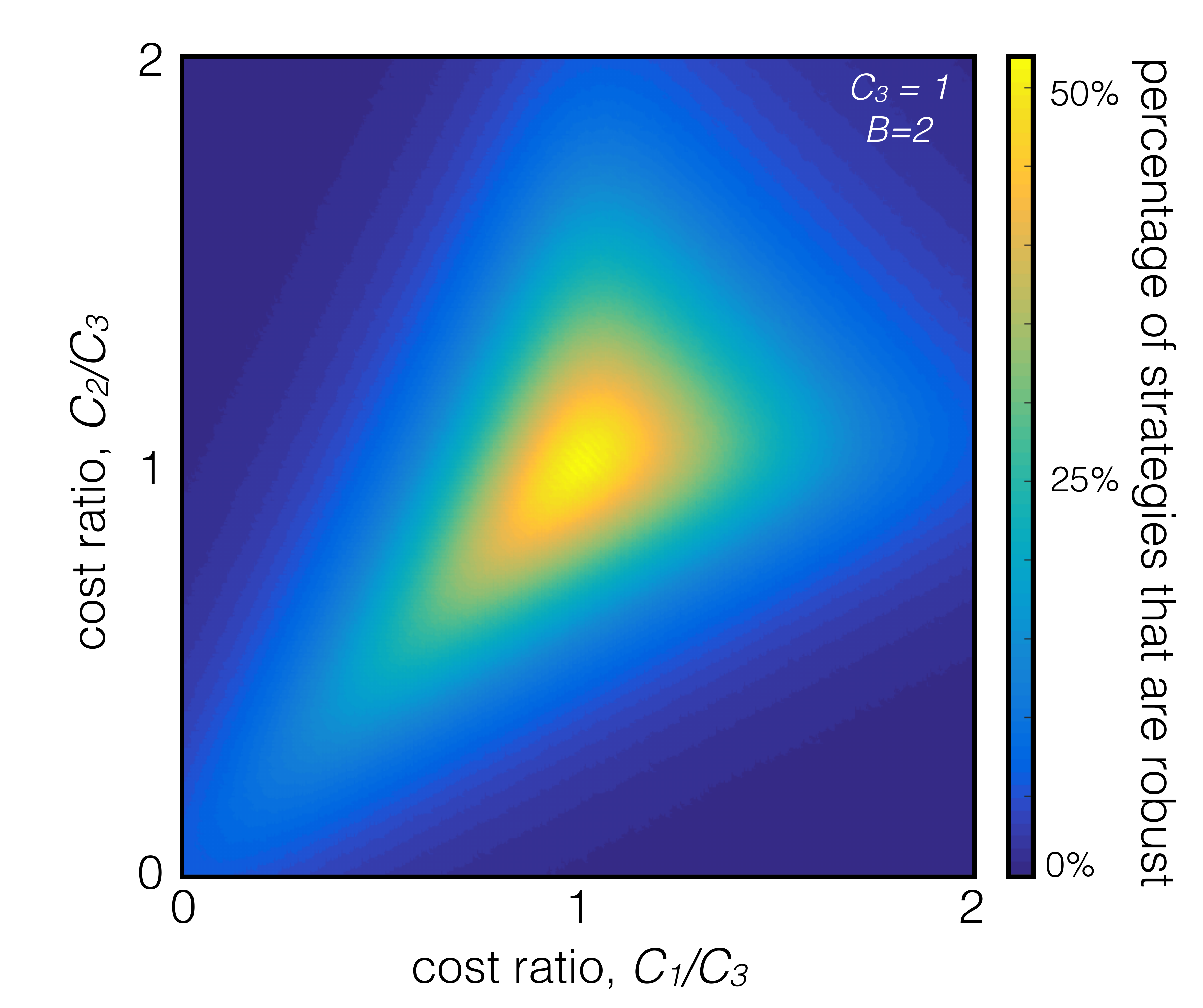}
\includegraphics[scale=0.22]{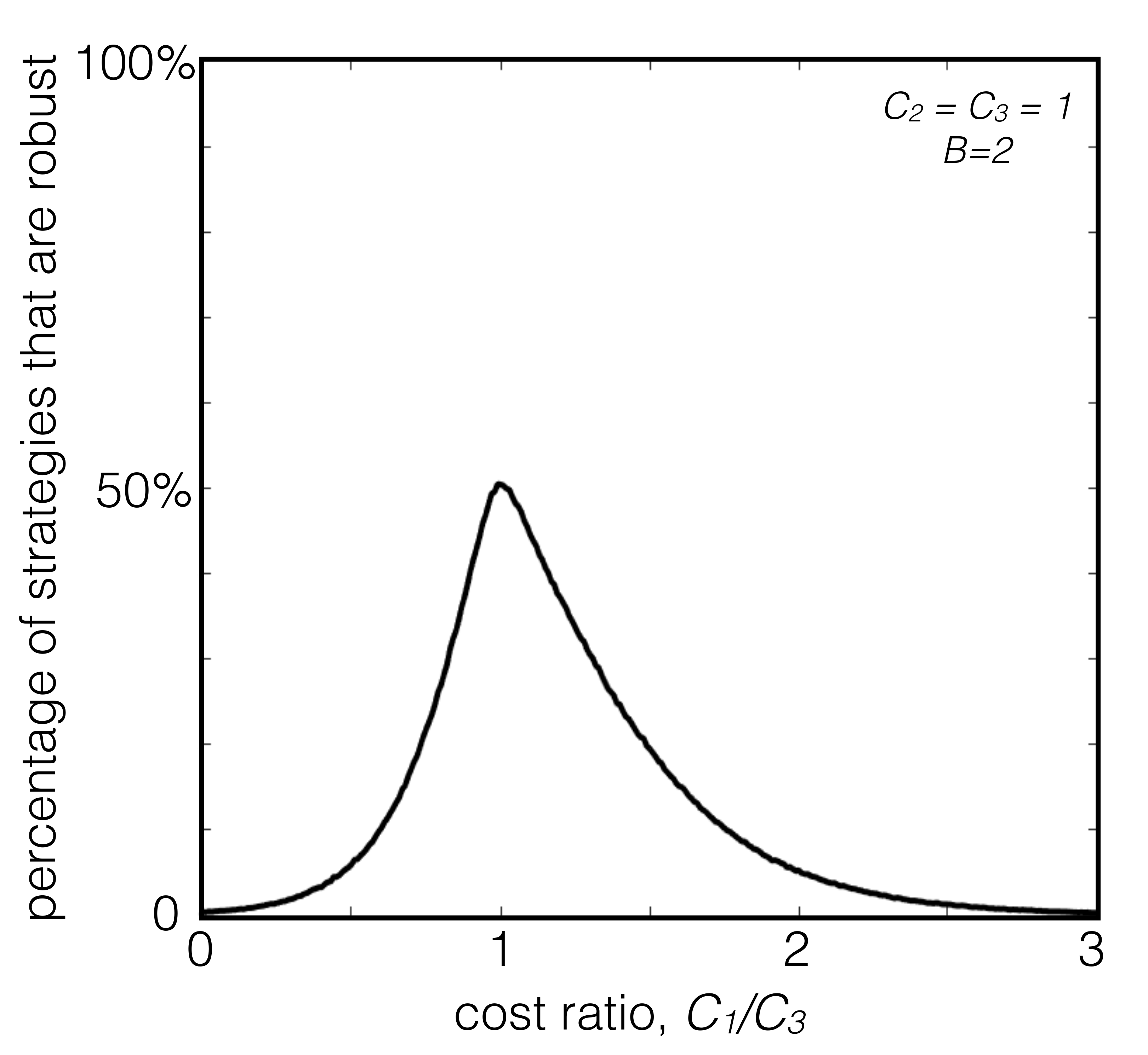} \caption{\small Can behavioral
diversity be maintained under non-transitive payoff structures? We considered a
rock-paper-scissors type game in which players could employ up to three different
behaviors, at a cost $C_1$, $C_2$ and $C_3$, in an attempt to obtain a fixed
benefit $B$.  The payoff structure was non-transitive so that action 1 dominates
action 2, action 2 dominates action 3, and action 3 dominates action 1. 
We determined whether a memory-1
strategy that employs all three behaviors can resist invasion by a player who
uses a single action exclusively (either 1, 2, or 3). 
(a) With fixed benefit $B=2$ and cost $C_3=1$ we
systematically varied costs $C_1$ and $C_2$, and  we calculated the percentage of
memory-1 strategies that could successfully maintain behavioral diversity.
Behavioral diversity can indeed be maintained for a wide range of costs. The
highest level of robust diverse strategies occurs in the symmetric case, when
$C_1=C_2=C_3$. But diverse behaviors are across a broad range of parameters
including, surprisingly, when both $C_1>B$ and $C_2>B$.  This is seen more
clearly in (b) which shows the percentage of robust strategies as a function of
$C_1$ with $C_2=C_3$.} \label{fig:figure} \end{figure*}

\section*{Discussion}

We have studied how the repertoire of behavioral options influences the prospects
for cooperation, and the maintenance of behavioral diversity, in evolving
populations. Our analysis has relied on the theory of iterated games and, in
particular, on a coordinate system we developed to describe strategies for
multi-choice games and their effects on long-term payoffs. In the context of
public goods games, we have shown that simple strategies that use only two levels
of investment can nonetheless stabilize cooperative behavior against arbitrarily
diverse mutant invaders, provided the simple strategy has sufficient opportunity
to punish defectors. More generally, a greater diversity of investment options can
either facilitate or hinder the evolution of cooperation, depending on the ratio
of public benefit produced to an individual's investment cost. We have applied the
same analytical framework to study more complicated multi-choice iterated games
with non-transitive payoffs, such as the rock-paper-scissors game. In this case,
behaviorally diverse strategies that employ multiple actions are often
evolutionary robust, even in a well-mixed population, and they can likewise invade
populations that lack diverse behaviors. Overall, the view emerges that simple
behavioral interactions are sometimes surprisingly robust against diverse
alternatives, and yet, in many circumstances, diverse behavior serves the mutual
benefit of a population and is a likely outcome of evolution.

Our results on the impact of multiple behavioral choices should be compared to
those of McAvoy \& Hauert \cite{McAvoy}, who studied ZD strategies in the
two-player donation game, with an arbitrary action space. Those authors
established that ZD strategies exist even in this general setting. They focused
especially on extortion strategies, whereby one player unilaterally sets the ratio
of scores against her opponent.  McAvoy \& Hauert found, remarkably, that
extortion strategies exists with support on only two actions, even against an
opponent who can choose from an uncountable number of actions. Our results form a
intriguing contrast to those of McAvoy \& Hauert. Instead of studying ZD
strategies and extortion in the classical context of two players, we have studied
all memory-1 strategies and the prospects for robust cooperation in a population
of $N>2$ players. We find that behaviorally depauperate strategies that rely on
only two actions can nonetheless sustain cooperation in a population facing
diverse invaders; and yet diversity can either hinder or facilitate cooperation,
depending upon the ratios of public benefit to individual cost.

We have analyzed the entire space of memory-1 strategies for iterated multi-choice
games. The purview of our analysis can be put in context by comparison to the yet
wider space of long-memory strategies, on the one hand, and the smaller space of
ZD strategies, on the other hand. As discussed here and elsewhere, strategies that
are evolutionary robust against the full space of memory-1 strategies are also
robust against all longer-memory strategies \cite{Press:2012fk,Pan:2015aa} (also
see Supporting Information), making this a natural strategy space to consider from
an evolutionary perspective. Nonetheless, memory can have an important impact on
the relative success of different types of robust strategies, by making them more
or less evolvable \cite{Stewart:2016aa}, or by allowing qualitatively different
types of decision-making via tagging or kin recognition \cite{Lee:2015aa}.
Conversely, it is important to consider the full space of memory-1 strategies in
the context of multi-choice games because, as we have shown, such games may
contain no ZD strategies at all, as in the case of iterated rock-papers-scissors.

It is perhaps unsurprising that games with non-transitive payoffs do not in
general admit the opportunity for one player to exert unilateral control over the
game's outcome via ZD strategies -- after all, a player cannot successfully extort
an opponent whose behavior is so diverse that it cannot be pinned down. Yet our
analysis also offers a novel perspective on the problem of diversity maintenance
in evolving populations.  One-shot rock-paper-scissors games have long been
studied in the context of evolutionary ecology as a system that cannot easily maintain
diversity without spatial structure or other exogenous population heterogeneity
\cite{Kerr:2002kx,Lively,Reichenbach:2007aa,Reichenbach:2007ab,Szolnoki:2014aa,Szczesny:2014aa,Bergstrom:2015aa}.
Here, by contrast, we have shown that behaviorally diverse strategies in the
iterated game can easily emerge and resist invasion by behaviorally depauperate
mutants, an observation which is relevant to behavioral interactions within a
single population and also to interactions between species.

Overall we have seen that, as players gain access to more behavioral choices,
either due to environmental shifts or evolutionary innovation, the dynamics of
social evolution can be profoundly altered. This view is reflected by empirical
studies, which have found that greater behavioral choice, via factors such as the
ability to communicate or signal to others, has a significant impact on the level
of cooperation in a group
\cite{Rand3,Rand:2012aa,Nowak:2006ys,Jordan:2016aa,Hauert:2007fk,NowakIndirect,Chen:2012aa}.
Moving forward, we must connect the insights drawn from complex behavioral and
evolutionary models of the type described here to empirical studies, where we can
now seek quantitative predictions for the dynamics of group behavior in real
populations. 

\bibliographystyle{pnas.bst}

\clearpage

\section*{Supporting Information}

\noindent In this supplement we first generalize the results of Press \& Dyson 2012 and Stewart \& Plotkin 2014 \cite{Stewart:2014aa,Press:2012fk} to the case of an infinitely iterated, $d$-choice, two-player game. We then apply those results to study evolutionary robustness of cooperation in a public goods game and maintenance of behavioral diversity in a rock-paper-scissors game.

\section*{Infinitely Iterated Multi-choice Games}

In this supplement, we generalize the results of \cite{Press:2012fk} to a game with an arbitrary number $d$ of pure strategies, which we refer to as different ``choices''. We start by repeating Press \& Dyson's argument for relating the payoffs for each player to a determinant.

The essential fact for their argument is that 1 is a simple (left) eigenvalue for an $n \times n$ Markov transition matrix $\bm{M}$. Recall that , for square matrices, the left and right eigenvalues are the same and have equal multiplicities (this is easily seen by observing that the characteristic equations for  $\bm{M}^{T}$ and  $\bm{M}$ are equal: $\det(\lambda \bm{I} - \bm{M}^{T}) = \det(\lambda \bm{I} - \bm{M})$). 
 
Now, 1 is always a left eigenvalue of any transition matrix - because the rows must sum to 1, the vector $\mathbbm{1}$ with all entries equal to 1 is a right eigenvector for the eigenvalue 1.  The only constraint to generalizing the result of \cite{Press:2012fk}  to more than two choices is that 1 must continue to be a simple eigenvalue \textit{i.e.}\, up-to-scalar multiples, the (left) eigenvector $\bm{v}$ such that $\bm{v}^{T}\bm{M} = \bm{v}^{T}$  must be unique (for the sake of concreteness, we will normalize $\bm{v}$ so that all it's entries sum to 1).  This is a consequence of the Perron-Frobenius Theorem, which says that if $\bm{M}$ is a non-negative (\ie all entries are non-negative), irreducible matrix, then the spectral radius of the matrix (here equal to 1) is a simple eigenvector.  We recall a matrix $\bm{A}$ is \textit{reducible} if there exists a permutation matrix $\bm{P}$ such that $\bm{P}\bm{A}\bm{P}^{T}$ is block upper triangular, and is \textit{irreducible} otherwise.  A more revealing equivalent expression for irreducibility is that  there exists $k$ such that $(A^{k})_{ij} > 0$ for all $i,j$, \ie the Markov chain has a positive probability of getting from state $i$ to state $j$ in finite time.  A two-player game is not necessarily irreducible, \eg the game in which player one always plays the choice they played in the previous round, and the eigenvector $\bm{v}$ need not be unique (in the aforementioned example, there are as many distinct eigenvectors as there are choices).  Nonetheless, reducible strategies are a lower dimensional subspace of all strategies, and are thus non-generic.


Now, suppose that $\bm{v}$ is the unique left eigenvector of $\bm{M}$ corresponding to the eigenvalue 1 and set $\bm{M}' \defn \bm{M} - \bm{I}$.  Then $\bm{v}$ is the unique vector such that $\bm{v}^{T}\bm{M}' = \bm{0}$, so 0 is an eigenvalue of $\bm{M}'$. Thus, $\det(\bm{M}') = 0$, and Cramer's rule tells us that 
\[
	\text{Adj}(\bm{M}')\bm{M}' = \det(\bm{M}') \bm{I} = 0,
\]
from which we conclude that every row of $\text{Adj}(\bm{M}')$ is a left eigenvector for the eigenvalue 0, and thus must be a scalar multiple of $\bm{v}$.  

Recall that, given an $n \times n$ matrix $\bm{A}$, the classical adjoint of $\bm{A}$, $\text{Adj}(\bm{A})$ is the matrix with entries equal to the cofactors of $\bm{A}$: 
\[
	\text{Adj}(\bm{A})_{ij} = (-1)^{i+j} \det(\bm{A}(i\vert j)),
\]
where $\bm{A}(i\vert j)$ is the $n-1 \times n-1$ matrix obtained by deleting the $i$\textsuperscript{th} row and $j$\textsuperscript{th} column of $\bm{A}$.  We also recall Laplace's cofactor expansion for the determinant:  for any choice of row $i$ or column $j$, we have
\[
	\det(\bm{A}) = \sum_{j=1}^{n} (-1)^{i+j} \det(\bm{A}(i\vert j)) a_{ij}  = \sum_{i=1}^{n} (-1)^{i+j} \det(\bm{A}(i\vert j)) a_{ij}.
\]
	
Now, in \cite{Press:2012fk}, the authors observe that if $\bm{f}$ is any column vector in $\mathbb{R}^{n}$ and $(\bm{A}\vert \bm{f})$ is the matrix obtained by replacing the $n$\textsuperscript{th} column of $\bm{A}$ with $\bm{f}$, then 
\[
	\det((\bm{A}\vert \bm{f})) 
	= \sum_{i=1}^{n} (-1)^{i+n} \det((\bm{A}\vert \bm{f})(i\vert n)) (\bm{A}\vert \bm{f})_{in} 
	= \sum_{i=1}^{n} (-1)^{i+n} \det(\bm{A}(i\vert d)) f_{i} 
	=  \sum_{i=1}^{n} \text{Adj}(\bm{A})_{in} f_{i} 
\]
(\nb $\bm{A}(i\vert n))$ is obtained by deleting the $n$\textsuperscript{th} column of $(\bm{A}\vert \bm{f})$, and thus is equal to $\bm{A}(i\vert n)$, whereas $(\bm{A}\vert \bm{f})_{in} = f_{i}$ by construction), and that this latter is the dot product of the $n$\textsuperscript{th} column of $\text{Adj}(\bm{A})$ with $\bm{f}$.  Now, as we have already observed, the $n$\textsuperscript{th} column of $\text{Adj}(\bm{M}')$ is $\alpha \bm{v}$, for some non-zero $\alpha$, so 
\[	
	\det((\bm{M}'\vert \bm{f})) = \alpha \bm{v} \cdot \bm{f}
\]
for arbitrary $\bm{f}$.  In particular, recalling that all entries of $\bm{v}$ sum to 1, we have
\[
	\det((\bm{M}'\vert \mathbbm{1})) = \alpha \bm{v} \cdot \mathbbm{1} = \alpha
\]
and thus 
\begin{equation}	
	\frac{\det((\bm{M}'\vert \bm{f}))}{\det((\bm{M}'\vert \mathbbm{1}))} = \bm{v} \cdot \bm{f}.
\end{equation}
Next, recall that $\det(\bm{A})$ is an alternating multilinear function of the columns of $\bm{A}$, so for arbitrary $m$, vectors $\bm{f}_{1},\cdots,\bm{f}_{m} \in \mathbb{R}^{n}$, and scalars $\alpha_{1},\ldots,\alpha_{m}$
\[
	\det\left(\left(\bm{A}\middle\vert \sum_{k=1}^{m} \alpha_{k} \bm{f}_{k}\right)\right) = \sum_{k=1}^{m} \alpha_{k} \det((\bm{A}\vert \bm{f}_{k})),
\]
and thus, 
\[
	\frac{\det\left(\left(\bm{M}'\vert \sum_{k=1}^{m} \alpha_{k} \bm{f}_{k}\right)\right)}{\det((\bm{M}'\vert \mathbbm{1}))} = \sum_{k=1}^{m} \alpha_{k} (\bm{v} \cdot \bm{f}_{k}).
 \]
Press \& Dyson then observe  that  player $i$'s payoff is $S_{i} \defn \bm{v} \cdot \bm{R}_{i}$, where $\bm{R}_{i}$ is the vector of payoffs received by player $i$ and $\bm{v}$ is the vector giving the equilibrium rate of different plays in an infinitely iterated game.  If there are 2 players, then 
\[
	\frac{\det\left(\left(\bm{M}'\vert \alpha_{1} \bm{R}_{1} + \alpha_{2} \bm{R}_{2} 
	+ \alpha_{3} \mathbbm{1} \right)\right)}{\det((\bm{M}'\vert \mathbbm{1}))} 
		= \alpha_{1} S_{12} + \alpha_{2} S_{21} + \alpha_{3}.
 \]

Now, to get the enforced relation
\[	
	\alpha_{1} S_{12} + \alpha_{2} S_{21} + \alpha_{3} = 0,
\]
Press \& Dyson use the alternating property of the determinant, namely that if any two columns are equal (or more generally, if there exists a subset of columns such that some linear combination of those columns is equal to one of the remaining columns) then the determinant is 0.

Thus, to generalize the result of \cite{Press:2012fk}, we need only verify that each of the two players can independently force the equality of at least two columns.  

The first step in doing this to recalling that for any matrix $\bm{A}$, $\det(\bm{A})$ is left unchanged by replacing any row or column by itself plus a linear combination of the other rows or columns, respectively.  Thus, if by such operations, we can transform $(\bm{M}'\vert \bm{f})$ to a matrix $\tilde{(\bm{M}'\vert \bm{f})}$ with one column that only depends on player $i$'s strategy, say $\bm{p}$, then player $i$ can enforce the linear relation (and, since $i$ is arbitrary, so can any other player) by setting a column that they control equal to 
\[
	  \alpha_{1} \bm{R}_{1} + \alpha_{2} \bm{R}_{2} + \alpha_{3} \mathbbm{1}.
\]

In what follows, we show that in the case of $d$ choices, which we label $0,\ldots,d-1$, the  transition matrix $\bm{M}$ is such that for an arbitrary vector $\bm{f} \in \mathbb{R}^{n}$ (here, $n = d^{2}$) $(\bm{M}'\vert \bm{f})$ has $d$ columns that are completely determined by player 1 and $d$ columns that are controlled by player 2.

We order the possible outcomes of play by the $d$-ary ordering.  That is to say, we denote the event where player 1 plays choice $j$ and player 2 strategy $k$ by $jk$, and order these events such that $jk$ is the $(d-1)j + k$\textsuperscript{th} possible outcome.  
Throughout this section, we will use $d=3$ as an example to clarify the discussion; in this case, we have possible plays
\[
	11, 12, 13, 21, 22, 23, 31, 32, 33
\]

Let $p^{i}_{jk}$ and $q^{i}_{kj}$  ($i=1,\ldots,d$, $j,k=1,\ldots,d$) denote the probabilities that player 1 and player 2 respectively use choice $i$ given that in the previous round player $1$ used choice $j$ and player $2$ used choice $k$ 
\[
	\sum_{j=1}^{k}\sum_{k=1}^{k}p^{i}_{jk} = 1
	\quad \text{and} \quad 
	\sum_{j=1}^{k}\sum_{k=1}^{k}q^{i}_{jk} = 1.	
\]
With this notation, the transition matrix $\bm{M}$ has entries 
\[
 	m_{i,jk} = p^{i}_{jk} q^{i}_{jk},
\]
which, for $d=3$ gives us
\[
	\bm{M} = \begin{bmatrix}
		p^1_{11}q^1_{11} & p^1_{11}q^2_{11} & p^1_{11}(1-q^1_{11}-q^2_{11}) & \cdots \\
		p^1_{12}q^1_{21} & p^1_{12}q^2_{21} &p^1_{12}(1-q^1_{21}-q^2_{21})& \cdots \\
		p^1_{13}q^1_{31} & p^1_{13}q^2_{31} & p^1_{13}(1-q^1_{31}-q^2_{31}) & \cdots \\
		p^1_{21}q^1_{12} & p^1_{21}q^2_{12}& p^1_{21}(1-q^1_{12}-q^2_{12}) & \cdots \\
		\vdots & \vdots & \vdots & \\
		p^1_{33}q^1_{33} & p^1_{33}q^2_{33} & p^1_{33}(1-q^1_{33}-q^2_{33}) & \cdots
	\end{bmatrix}
\]

Next, $\bm{M}'$ has entry $m_{i,j}' = m_{i,j} - \delta_{i,j}$, where $\delta_{i,j}$ is Kronecker's delta function.  Again, for $d=3$, this gives
\[
	\bm{M}' =  \begin{bmatrix}
		p^1_{11}q^1_{11} -1& p^1_{11}q^2_{11} & p^1_{11}(1-q^1_{11}-q^2_{11}) & \cdots \\
		p^1_{12}q^1_{21} & p^1_{12}q^2_{21}-1 &p^1_{12}(1-q^1_{21}-q^2_{21})& \cdots \\
		p^1_{13}q^1_{31} & p^1_{13}q^2_{31} & p^1_{13}(1-q^1_{31}-q^2_{31})-1 & \cdots \\
		p^1_{21}q^1_{12} & p^1_{21}q^2_{12}& p^1_{21}(1-q^1_{12}-q^2_{12}) & \cdots \\
		\vdots & \vdots & \vdots & \\
		p^1_{33}q^1_{33} & p^1_{33}q^2_{33} & p^1_{33}(1-q^1_{33}-q^2_{33}) & \cdots
	\end{bmatrix}
\]
Finally, the row corresponding to the plays $jk$ of $(\bm{M}'\vert \bm{f})$ has entries
\begin{equation*}
	p^{1}_{jk}q^{1}_{kj},\ldots,p^{1}_{jk}q^{d}_{kj} ,
	p^{2}_{jk}q^{1}_{kj},\ldots,p^{j}_{jk}q^{k}_{kj} - 1,\ldots,
	p^{d}_{jk}q^{1}_{kj} ,\ldots,p^{d}_{jk}q^{d-1}_{kj}, f_{jk},
\end{equation*}
and continuing to illustrate this with $d=3$, we have
\[
	(\bm{M}'\vert \bm{f}) = \begin{bmatrix}
		p^1_{11}q^1_{11} -1& p^1_{11}q^2_{11} & p^1_{11}(1-q^1_{11}-q^2_{11}) & \cdots,f_{11} \\
		p^1_{12}q^1_{21} & p^1_{12}q^2_{21}-1 &p^1_{12}(1-q^1_{21}-q^2_{21})& \cdots,f_{12} \\
		p^1_{13}q^1_{31} & p^1_{13}q^2_{31} & p^1_{13}(1-q^1_{31}-q^2_{31})-1 & \cdots,f_{13} \\
		p^1_{21}q^1_{12} & p^1_{21}q^2_{12}& p^1_{21}(1-q^1_{12}-q^2_{12}) & \cdots,f_{21} \\
		\vdots & \vdots & \vdots & \\
		p^1_{33}q^1_{33} & p^1_{33}q^2_{33} & p^1_{33}(1-q^1_{33}-q^2_{33}) & \cdots,f_{33}
	\end{bmatrix}
\]

Thus, the sum of the first $d$ entries of the $jk$\textsuperscript{th} row of $(\bm{M}'\vert \bm{f})$ is 
\begin{equation*}
 p^{1}_{jk}q^{1}_{kj} + \cdots + p^{1}_{jk}q^{d}_{kj}
	=\begin{cases}
		 p^{1}_{jk}-1 & \text{if $j=1$}\\
		 p^{1}_{jk} & \text{otherwise}
	\end{cases}
\end{equation*}
Similarly for the second $d$ entries, and so on.  Thus, if for each $a=0,\ldots,d-1$, we replace the $ad$\textsuperscript{th} column by the sum of columns $ad, ad + 1, \ldots,  ad +  d-1$, a transformation that leaves $\det((\bm{M}'\vert \bm{f}))$ unchanged, the resulting matrix has a $ad$\textsuperscript{th} column with $jk$\textsuperscript{th} entry 
\[
	\begin{cases}
		 p^{a+1}_{jk}-1 & \text{if $j=a+1$}\\
		 p^{a+1}_{jk} & \text{otherwise}
	\end{cases}
\]	
\ie the $a d$\textsuperscript{th} column depends only on player 1, and player 1 controls $d$ columns, one for each available choice.  Proceeding similarly, we see that player 2 also controls exactly $d$ columns.

To see this concretely, for $d=3$, if we replace the third column of $(\bm{M}'\vert \bm{f})$ by the third column plus the first and the second (which preserves the determinant), we get 
\begin{equation}
	(\bm{M}'\vert \bm{f}) = \begin{bmatrix}
		p^1_{11}q^1_{11} -1& p^1_{11}q^2_{11} & p^1_{11}-1 & \cdots,f_{11} \\
		p^1_{12}q^1_{21} & p^1_{12}q^2_{21}-1 &p^1_{12}-1& \cdots,f_{12} \\
		p^1_{13}q^1_{31} & p^1_{13}q^2_{31} & p^1_{13}-1 & \cdots,f_{13} \\
		p^1_{21}q^1_{12} & p^1_{21}q^2_{12}& p^1_{21}& \cdots,f_{21} \\
		\vdots & \vdots & \vdots & \\
		p^1_{33}q^1_{33} & p^1_{33}q^2_{33} & p^1_{33} & \cdots,f_{33}
	\end{bmatrix}
\end{equation}
Thus, player 1 controls the third column of $\tilde{(\bm{M}'\vert \bm{f})}$ with their probabilities of playing choice 1.  Similarly replacing column 6 with the sum of columns 4, 5, and 6, we get a new column 6 with entries 
\[
	p^2_{11}, p^2_{12}, p^2_{13}, p^2_{21} - 1, p^2_{22} - 1, p^2_{23} - 1, p^2_{31} , p^2_{32}, p^2_{33}
\]
to conclude that player 1 controls 2 columns.  

\subsection*{Memory in multi-choice games}

Appendix A of \cite{Press:2012fk} tells us that if player 1 has memory $m_{1}$ and player 2 has memory $m_{2} > m_{1}$, then for any strategy played by player 2, there is a memory $m_{1}$ strategy that will yield the same expected payoff, which should be qualified by clarifying that the expected payoff refers to expectation with respect to all possible histories (as opposed to, say, expectation conditional on a given history of play).  Let $\mathcal{H}_{n}$ denote the history of plays up until the $n$\textsuperscript{th} round, and let $S_{1}(n), S_{2}(n)$ denote the strategy played by player 1 and 2 respectively in the $n$\textsuperscript{th} round; then player $i$ has memory $m_{i}$ is the statement that 
\[ 
	\mathbb{E}[S_{i}(n) = s \vert \mathcal{H}_{n}]
	= \mathbb{E}\left[S_{i}(n) = s 
		\vert (S_{1}(n-1),S_{2}(n-1)),\ldots,(S_{1}(n-m_{i}),S_{2}(n-m_{i}))\right]
\]
Now, let $\tilde{S}_{2}$ be a random variable such that
\begin{multline*}
	\mathbb{P}\left(\tilde{S}_{2}(n) = s
		\middle\vert (S_{1}(n-1),S_{2}(n-1)),\ldots,(S_{1}(n-m_{1}),S_{2}(n-m_{1}))\right)\\
	= \mathbb{E}\left[\mathbb{P}\left(S_{2}(n) = s
		\middle\vert (S_{1}(n-1),S_{2}(n-1)),\ldots,(S_{1}(n-m_{2}),S_{2}(n-m_{2}))\right)\right],
\end{multline*}
where the expectation is over the outcomes of the plays $(S_{1}(n-m_{1}),S_{2}(n-m_{1})),\ldots,(S_{1}(n-m_{2}),S_{2}(n-m_{2}))$.  Then $\tilde{S}_{2}$ is a memory $m_{1}$ strategy and it is shown in 
\cite{Press:2012fk} that player 1 has the same payoff playing against the new player $\tilde{S}_{2}$ as against the original opponent playing $S_{2}$.  Since the Nash equilibrium depends only on the expected payoff, this tells us that we may equally well determine the Nash equilibrium by playing against the shorter memory player.

\section*{Coordinate system for memory-1 strategies in multi-choice games}
Just as in the case of two-choice games, we can use Eqs. 1 and 2 to construct a coordinate system for the space of memory-1 strategies. Consider a $d$-choice, two-player game with strategy $(\mathbf{p}^1,\mathbf{p}^2,\ .\ .\ .,\mathbf{p}^d)$ where each $\mathbf{p}^i$ is a vector of $d^2$ probabilities, each corresponding to the probability that a player makes choice $i$ in the next round given the outcome of the preceding round. By definition we must have $\sum_i^kp^i_j=1, \forall j\in \mathcal{D}$ where $\mathcal{D}$ is the set of possible choices in the game. In order to construct an alternate coordinate system we must choose $d^2$ vectors that form a basis $\mathbb{R}^{d^2}$. To do this we choose $d(d+1)/2$ vectors that have entry $1$ at the $i$th and $j$th position for all pairs $i,j$ and entry zero otherwise. We also choose $d(d-1)/2$ vectors that have entry 1 at the $i$th and entry $-1$ at the $j$th for all paris $i,j$, where we adopt the convention that the first entry is positive. The new coordinate system is the $\left\{\Lambda^+_{11},\Lambda^+_{12},. \ . \ .,\Lambda^+_{1d},. \ . \ .,\Lambda^+_{dd},\Lambda^-_{1d},. \ . \ .,\Lambda^-_{1d},. \ . \ .,\Lambda^-_{d-1d}\right\}$ and we have in the case $d=3$

\begin{equation}
\det\left[ \begin{array}{ccccccccc}
1& 0 & 0 & 0 & 0 & 0 & 0 & 0 &0\\
0 & 1 & 0 & 0 & 0 & 0 & 1 & 0 &0\\
0 & 0 & 1  &  0 & 0 & 0 & 0 & 1& 0 \\
0 & 1 & 0 & 0 & 0 & 0 & -1 & 0 & 0\\
0 & 0 & 0 & 1 & 0 & 0 & 0 & 0 & 0\\
0 & 0 & 0 & 0 & 1 & 0 & 0 & 0 & 1\\
0 & 0 & 1 & 0 & 0 & 0 & 0 & -1& 0 \\
0 & 0 & 0 & 0 & 1 & 0 & 0 & 0 & -1\\
0 & 0 & 0 & 0 & 0 & 1 & 0 & 0 & 0\\
\end{array} \right]=-8\end{equation}
\\
which is a basis $\mathbb{R}^{9}$ as required. From Eqs. 1 and 2 we then end up with

\begin{equation}
\sum_{i=1}^d\left(\Lambda^+_{ii}v_{ii}+\sum_{j=i+1}^d\Lambda^+_{ij}(v_{ij}+v_{ji})+\Lambda^-_{ij}(v_{ij}-v_{ji})\right)=0
\end{equation}
\\
where $v_{ij}$ is the equilibrium rate of the play $ij$, with the focal player's move is listed first.
Now let the expected payoff to a focal player $X$ and her opponent $Y$ to be $S_{xy}$ and $S_{yx}$ respectively. By definition these satisfy:

\begin{equation}
S_{xy}+S_{yx}=\sum_{i=1}^d\left(2R_{ii}v_{ii}+\sum_{j=i+1}^d(R_{ij}+R_{ji})(v_{ij}+v_{ji})\right)
\end{equation}
\\
and

\begin{equation}
S_{xy}-S_{yx}=\sum_{i=1}^d\sum_{j=i}^d(R_{ij}-R_{ji})(v_{ij}-v_{ji})
\end{equation}
\\
where $R_{ij}$ is the payoff to the focal player in a given round in which she played $i$ and her opponent $j$. Note also that

\begin{equation}
\sum_{i=1}^d\sum_{j=1}^kv_{ij}=1
\end{equation}
\\
be definition.
If we now set

\begin{equation}
	\Lambda^+_{ij} = \frac{\phi-\chi}{2}(R_{ij}+R_{ji})-(\phi-\chi)\kappa+\lambda^+_{ij}\\
\end{equation}
\\
and

\begin{equation}
	\Lambda^+_{ij} = -\frac{\phi+\chi}{2}(R_{ij}-R_{ji})+\lambda^-_{ij}\\
\end{equation}
\\
and define

\[
\lambda_{ij}=\lambda^+_{ij}+\lambda^-_{ij}
\]
\\
and

\[
\lambda_{ji}=\lambda^+_{ij}-\lambda^-_{ij}
\]
\\
for all $j>i$, we can combine Eqs.4-9 to recover the following relationship:

\begin{equation}
\phi S_{yx}-\chi S_{xy}-(\phi-\chi)\kappa+\sum_{i=1}^d\sum_{j=1}^d\lambda_{ij}v_{ij}=0
\end{equation}
\\
Notice that we now have three extraneous parameters. In general a convenient choice is $\lambda_{11}=\lambda_{dd}=0$ and $\lambda_{1d}=\lambda_{d1}$, however more convenient choices can be made depending on the payoff structure of the game being considered.
Under this coordinate system, for a game with $d=3$ we end up with

\begin{eqnarray*}
p^1_{11}&=&1-\left(\phi^1R_{11}-\chi^1R_{11}-(\phi^1-\chi^1)\kappa^1\right)\\
p^1_{12}&=&1-\left(\phi^1R_{21}-\chi^1R_{12}-(\phi^1-\chi^1)\kappa^1+\lambda_{12}\right)\\
p^1_{13}&=&1-\left(\phi^1R_{31}-\chi^1R_{13}-(\phi^1-\chi^1)\kappa^1+\lambda_{31}\right)\\
p^1_{21}&=&-\left(\phi^1R_{12}-\chi^1R_{21}-(\phi^1-\chi^1)\kappa^1+\lambda_{21}\right)\\
p^1_{22}&=&-\left(\phi^1R_{22}-\chi^1R_{22}-(\phi^1-\chi^1)\kappa^1+\lambda_{22}\right)\\
p^1_{23}&=&-\left(\phi^1R_{32}-\chi^1R_{23}-(\phi^1-\chi^1)\kappa^1+\lambda_{23}\right)\\
p^1_{31}&=&-\left(\phi^1R_{13}-\chi^1R_{31}-(\phi^1-\chi^1)\kappa^1+\lambda_{31}\right)\\
p^1_{32}&=&-\left(\phi^1R_{23}-\chi^1R_{32}-(\phi^1-\chi^1)\kappa^1+\lambda_{32}\right)\\
p^1_{33}&=&-\left(\phi^1R_{33}-\chi^1R_{33}-(\phi^1-\chi^1)\kappa^1\right)\\
\end{eqnarray*}
\\
where we have used the superscript 1 to indicate that this is the probability of choosing to play 1. Clearly the same argument holds for choices 2 and 3, with the caveat that $\sum_i^kp^i_j=1, \forall j\in \mathcal{D}$.

We now use this coordinate system to analyse two multi-choice cases of particular interest: two-choice strategies playing against multi-choice invaders in a public goods game, and multi-choice strategies playing against single choice invaders in a rock-paper scissors game.

\section*{Robust strategies in multi-choice public goods games}

We now turn our attention to a multi-choice public goods game, in which a pair of players who invest $C_j$ and $C_k$ respectively in a given round of play generate a total benefit $B_{jk}$ such that

\[
R_{jk}=B_{jk}/2-C_j
\]
\\
We are interested in whether a two-choice strategy can be evolutionary robust against an invader who can vary his investment level in an arbitrary way. Thus we assume a focal strategy that can invest either $C_1$ or $C_2$. We assume $\lambda_{11}=\lambda_{22}=0$ and $\lambda_{12}=\lambda_{21}$. When faced with an opponent who plays with $d$ investment levels, the two-choice player may in general have $2d$ probabilities for cooperation

\begin{eqnarray*}
p^1_{11}&=&1-\left((\phi -\chi )(B_{11}/2-\kappa)-\phi C_1+\chi C_1\right)\\
p^1_{12}&=&1-\left((\phi -\chi )(B_{12}/2-\kappa)-\phi C_2+\chi C_1+\lambda_{12}\right)\\
p^1_{13}&=&1-\left((\phi -\chi )(B_{13}/2-\kappa)-\phi C_3+\chi C_1+\lambda_{13}\right)\\
\vdots\\
p^1_{1d}&=&1-\left((\phi -\chi )(B_{1d}/2-\kappa)-\phi C_d+\chi C_1+\lambda_{1d}\right)\\
p^1_{21}&=&-\left((\phi -\chi )(B_{12}/2-\kappa)-\phi C_1+\chi C_2+\lambda_{12}\right)\\
p^1_{22}&=&-\left((\phi -\chi )(B_{22}/2-\kappa)-\phi C_2+\chi C_2\right)\\
p^1_{23}&=&-\left((\phi -\chi )(B_{23}/2-\kappa)-\phi C_3+\chi C_2+\lambda_{23}\right)\\
\vdots\\
p^1_{2d}&=&-\left((\phi -\chi )(B_{2d}/2-\kappa)-\phi C_d+\chi C_2+\lambda_{2d}\right)\\
\end{eqnarray*}
\\
where $p^2_{jk}=1-p^1_{jk}$. The resulting relationship between players' scores is given by

\begin{equation}
\phi S_{yx}-\chi S_{xy}-(\phi-\chi)\kappa+\lambda_{12}(v_{12}+v_{21})+\sum_{j=3}^d(\lambda_{1j}v_{1j}+\lambda_{2j}v_{2j})=0
\end{equation}
\\
We can observe immediately that the first four terms of Eq. 11 corresponds to the type of two-choice games that have been studied extensively elsewhere.

Looking at the sum and difference between players' scores in this game we find

\begin{equation}
S_{xy}+S_{yx}=(B_{11}-2C_1)v_{11}+(B_{22}-2C_2)v_{22}+(B_{12}-C_1-C_2)(v_{12}+v_{12})+\sum_{j=3}^d(B_{1j}-C_1-C_j)v_{1j}+(B_{2j}-C_2-C_j)v_{2j}
\end{equation}
\\
and

\begin{equation}
S_{xy}-S_{yx}=(C_2-C_1)(v_{12}-v_{21})+\sum_{j=3}^d(C_j-C_1)v_{1j}+(C_j-C_2)v_{2j}
\end{equation}
\\
Now let us focus on a resident, two-choice strategy who can invest either $C_1$ or $C_2$ where $C_1>C_2$, and which stabalizes cooperation investment at $C_1$ when resident in a population, i.e such that $\kappa=B_{11}/2-C_1$. We have bounds on players scores of

\begin{eqnarray}
\nonumber S_{xy}+S_{yx}\leq(B_{11}-2C_1)+(B_{12}+C_1-C_2-B_{11})(v_{12}+v_{12})\\
+\sum_{j=3}^d(B_{1j}+C_1-C_j-B_{11})v_{1j}+(B_{2j}-C_2-C_j-B_{11}+2C_1)v_{2j}
\end{eqnarray}
\\
which becomes an equality when $v_{22}=0$, and

\begin{eqnarray}
\nonumber S_{xy}+S_{yx}\geq(B_{22}-2C_2)+(B_{12}-C_1+C_2-B_{22})(v_{12}+v_{12})\\
+\sum_{j=3}^d(B_{1j}-C_1-C_j-B_{22}+2C_2)v_{1j}+(B_{2j}+C_2-C_j-B_{22})v_{2j}
\end{eqnarray}
\\
which becomes an equality when $v_{11}=0$, and

\begin{equation}
S_{xy}-S_{yx}\geq-(C_1-C_2)(v_{12}+v_{21})+\sum_{j=3}^d(C_j-C_1)v_{1j}+(C_j-C_2)v_{2j}
\end{equation}
\\
which becomes an equality when an opponent never invests $C_2$ and

\begin{equation}
S_{xy}-S_{yx}\leq(C_1-C_2)(v_{12}+v_{21})+\sum_{j=3}^d(C_j-C_1)v_{1j}+(C_j-C_2)v_{2j}
\end{equation}
\\
which becomes an equality when an opponent never invests $C_1$.

In order for a rare mutant $Y$ to invade a population with a resident $X$ we must have

\begin{equation}
S_{yx}>\frac{N-2}{N-1}(B_{11}/2-C_1)+\frac{1}{N-1}S_{xy}
\end{equation}
\\
Combining this with Eq. 11 we then get

\begin{equation}
\left(\chi-\phi\frac{1}{N-1}\right)(S_{xy}-(B_{11}/2-C_1))>\lambda_{12}(v_{12}+v_{21})+\sum_{j=3}^d(\lambda_{1j}v_{1j}+\lambda_{2j}v_{2j})
\end{equation}
\\
Combining this with Eq14. and Eq. 16 we then get two conditions for evolutionary robustness, firstly
\begin{eqnarray}
\nonumber \frac{N}{N-1}\left(\lambda_{12}(v_{12}+v_{21})+\sum_{j=3}^d(\lambda_{1j}v_{1j}+\lambda_{2j}v_{2j})\right)>\\
\nonumber\left(\chi-\phi\frac{1}{N-1}\right)\Bigg[(B_{12}+C_1-C_2-B_{11})(v_{12}+v_{12})\\
+\sum_{j=3}^d(B_{1j}+C_1-C_j-B_{11})v_{1j}+(B_{2j}-C_2-C_j-B_{11}+2C_1)v_{2j}\Bigg]
\end{eqnarray}
\\
which means that we must have

\begin{equation}
\frac{N}{N-1}\lambda_{ij}>-\left(\chi-\phi\frac{1}{N-1}\right)(B_{11}-2C_1-B_{ij}+C_i+C_j)
\end{equation}
\\
We also get

\begin{eqnarray}
\nonumber\frac{N-2}{N-1}\lambda_{12}(v_{12}+v_{21})+\frac{N-2}{N-1}\sum_{j=3}^d(\lambda_{1j}v_{1j}+\lambda_{2j}v_{2j})\\
>-\left(\chi-\phi\frac{1}{N-1}\right)\left[(C_1-C_2)(v_{12}+v_{21})-\sum_{j=3}^d(C_j-C_1)v_{1j}+(C_j-C_2)v_{2j}\right]
\end{eqnarray}
\\
which means we must have

\begin{equation}
\frac{N-2}{N-1}\lambda_{ij}>\left(\chi-\phi\frac{1}{N-1}\right)(C_j-C_i), \ \ \forall j>2
\end{equation}
\\
This second equation is always hardest to satisfy when $C_j$  is minimized. For the former condition, we assume $B_{ij}=r(C_i+C_j)^\alpha$ to get
\[
\frac{N}{N-1}\lambda_{ij}>\left(\chi-\phi\frac{1}{N-1}\right)(r(2C_1)^\alpha-2C_1-r(C_i+C_j)^\alpha+(C_i+C_j))
\]
\\
which is hardest to satisfy when the right hand side is maximized. When this occurs depends in general on the choice of $\alpha$, but if $\alpha=1$
this condition is also hardest to satisfy when $C_j=0$.
Thus we have:

\begin{eqnarray}
\nonumber \frac{N}{N-1}\lambda_{10}&>&-\left(\chi-\phi\frac{1}{N-1}\right)(r-1)C_1
\\
\nonumber \frac{N}{N-1}\lambda_{20}&>&-\left(\chi-\phi\frac{1}{N-1}\right)((r-1)(2C_1-C_2))
\\
\nonumber \frac{N-2}{N-1}\lambda_{10}&>&\left(\chi-\phi\frac{1}{N-1}\right)C_1
\\
\nonumber \frac{N-2}{N-1}\lambda_{20}&>&\left(\chi-\phi\frac{1}{N-1}\right)C_2
\\
\end{eqnarray}
\\
as our conditions for a two-choice strategy to be robust.
We can also convert Eq. 20-23 back to the original coordinate system to give the following robustness conditions

\begin{align}
\nonumber  &\mathcal{C}^{d}_{s}=\Bigg\{(p_{11},p_{12},\ldots,p_{1d},p_{21},p_{22},\ldots,p_{2d})\bigg|p_{11}=1,\\
\nonumber &p_{1j}<1-\frac{N-2}{N}(1-p_{12}+p_{21})\frac{1-c^*}{1-c}\left[\frac{N-1}{N-2}-\frac{r}{2}\right], \\
\nonumber  &p_{2j}<
\frac{N-2}{N}(1-p_{12}+p_{21})\left[\frac{r}{2}-\left(\frac{N-1}{N-2}-\frac{r}{2}\right)\frac{1-c^*}{1-c}+\frac{1}{N-2}\right],\\
\nonumber &p_{1j}<1-\frac{p_{22}}{r-1}\frac{1-c^*}{1-c}\left[\frac{N-1}{N-2}-\frac{r}{2}\right]\\
\nonumber &p_{2j}<\frac{p_{22}}{r-1}\left[\frac{r}{2}-\left(\frac{N-1}{N-2}-\frac{r}{2}\right)\frac{1-c^*}{1-c}+\frac{1}{N-2}\right]
\Bigg\},
\end{align}
\begin{equation}
\end{equation}
\\
which is Eq. 3 of the main text. Finally in order for a strategy to be robust it must be viable, in addition to satisfying Eq. 25,
which leaves us with the condition

\begin{equation}
\frac{r-1}{\frac{r}{2}+\frac{1}{N-2}}>\frac{C_2}{C_1}
\end{equation}
\\
which must be satisfied in order for a robust two-choice strategy to exist.

\section*{Games with non-transitive payoff structures}
We now consider the rock-paper-scissors game, which is a three-choice, non-transitive game. We assume a payoff structure $R_{13}=B-C_1$, $R_{21}=B-C_2$, $R_{32}=B-C_3$, $R_{31}=-C_3$, $R_{12}=-C_1$ and $R_{23}=-C_2$ which gives a non-transitive relationship between the choices 1=rock, 2=paper and 3=scissors. We assume that when two players make the same choice they receive equal payoff: $R_{11}=B/2-C_1$, $R_{22}=B/2-C_1$ and $R_{33}=B/2-C_1$. In the alternate coordinate system a strategy
is written as

\begin{eqnarray*}
p^1_{11}&=&1-(\phi^1-\chi^1)\left(B/2-C_1-\kappa^1\right)\\
p^1_{12}&=&1-\left(\phi^1(B-C_2)+\chi^1C_1-(\phi^1-\chi^1)\kappa^1\right)\\
p^1_{13}&=&1+\left(\phi^1C_3+\chi^1(B-C_1)+(\phi^1-\chi^1)\kappa^1\right)\\
p^1_{21}&=&\lambda^1_{21}+\left(\phi^1C_2+\chi^1(B-C_1)+(\phi^1-\chi^1)\kappa^1\right)\\
p^1_{22}&=&\lambda^1_{22}-(\phi^1-\chi^1)\left(B/2-C_2-\kappa^1\right)\\
p^1_{23}&=&\lambda^1_{23}-\left(\phi^1(B-C_3)+\chi^1C_2-(\phi^1-\chi^1)\kappa^1\right)\\
p^1_{31}&=&\lambda^1_{31}-\left(\phi^1(B-C_1)+\chi^1C_3-(\phi^1-\chi^1)\kappa^1\right)\\
p^1_{32}&=&\lambda^1_{32}+\left(\phi^1C_2+\chi^1(B-C_3)+(\phi^1-\chi^1)\kappa^1\right)\\
p^1_{33}&=&\lambda^1_{33}-(\phi^1-\chi^1)\left(B/2-C_3-\kappa^1\right)\\
\end{eqnarray*}
\\
and

\begin{eqnarray*}
p^2_{11}&=&\lambda^2_{11}-(\phi^2-\chi^2)\left(B/2-C_1-\kappa^2\right)\\
p^2_{12}&=&\lambda^2_{12}-\left(\phi^2(B-C_2)+\chi^2C_1-(\phi^2-\chi^2)\kappa^2\right)\\
p^2_{13}&=&\lambda^2_{13}+\left(\phi^2C_3+\chi^2(B-C_1)+(\phi^2-\chi^2)\kappa^2\right)\\
p^2_{21}&=&1+\left(\phi^2C_2+\chi^2(B-C_1)+(\phi^2-\chi^2)\kappa^2\right)\\
p^2_{22}&=&1-(\phi^2-\chi^2)\left(B/2-C_2-\kappa^2\right)\\
p^2_{23}&=&1-\left(\phi^2(B-C_3)+\chi^2C_2-(\phi^2-\chi^2)\kappa^2\right)\\
p^2_{31}&=&\lambda^2_{31}-\left(\phi^2(B-C_1)+\chi^2C_3-(\phi^2-\chi^2)\kappa^2\right)\\
p^2_{32}&=&\lambda^2_{32}+\left(\phi^2C_2+\chi^2(B-C_3)+(\phi^2-\chi^2)\kappa^2\right)\\
p^2_{33}&=&\lambda^2_{33}-(\phi^2-\chi^2)\left(B/2-C_3-\kappa^2\right)\\
\end{eqnarray*}
\\
where we set $\lambda=0$ for the case where a player uses the same move as she played in the preceding round.
If we consider the symmetrical case $C_1=C_2=C_3$ we can set

\begin{eqnarray*}
p^o_{o}&=&1-(\phi-\chi)\left(B/2-C-\kappa\right)\\
p^-_{-}&=&1-\left(\phi(B-C)+\chi C-(\phi-\chi)\kappa\right)\\
p^+_{+}&=&1+\left(\phi C+\chi(B-C)+(\phi-\chi)\kappa\right)\\
p^o_{+}&=&\lambda^o_{+}+\left(\phi C+\chi(B-C)+(\phi-\chi)\kappa\right)\\
p^-_{o}&=&\lambda^-_{o}-(\phi-\chi)\left(B/2-C-\kappa\right)\\
p^+_{-}&=&\lambda^+_{-}-\left(\phi(B-C)+\chi C-(\phi-\chi)\kappa\right)\\
p^o_{-}&=&\lambda^o_{-}-\left(\phi(B-C)+\chi C-(\phi-\chi)\kappa\right)\\
p^-_{+}&=&\lambda^-_{+}+\left(\phi C+\chi(B-C)+(\phi-\chi)\kappa\right)\\
p^+_{o}&=&\lambda^+_{o}-(\phi-\chi)\left(B/2-C-\kappa\right)\\
\end{eqnarray*}
\\
where subscript indicates the outcome of the preceding round -- win (+), lose (-) or draw (o) and the superscript refers to the choice to switch to the move that would have resulted in that outcome in the preceding round.
Note also that by definition $p^+_{o}+p^-_o+p^o_o=1$ etc so that the following must hold:

\begin{eqnarray}
\nonumber  \lambda^-_{o}+\lambda^+_{o}&=&3(\phi-\chi)\left(B/2-C-\kappa\right)\\
\nonumber  \lambda^o_{+}+\lambda^-_{+}&=&-3\left(\phi C+\chi(B-C)+(\phi-\chi)\kappa\right)\\
\lambda^+_{-}+\lambda^o_{-}&=&3\left(\phi(B-C)+\chi C-(\phi-\chi)\kappa\right)
\end{eqnarray}
\\
Against an opponent who only plays rock=1, the following relationships between players scores must hold

\begin{eqnarray}
\nonumber \phi S_{yx}-\chi S_{xy}-(\phi-\chi)\kappa+\lambda^o_{+}v_{21}+\lambda^o_{-}v_{31}=0\\
\nonumber \phi S_{yx}-\chi S_{xy}-(\phi-\chi)\kappa+\lambda^+_{o}v_{11}+\lambda^+_{-}v_{31}=0\\
\nonumber \phi S_{yx}-\chi S_{xy}-(\phi-\chi)\kappa+\lambda^-_{o}v_{11}+\lambda^-_{+}v_{21}=0\\
\end{eqnarray}
\\
with equivalent equalities for invaders who only play paper or scissors, which we can ignore due to the assumed symmetry of the problem.

Finally, note that in the totally symmetrical game the sum of both players longterm average payoffs is constant:

\begin{equation}
S_{xy}+S_{yx}=B-2C
\end{equation}
\\
and in order for a mutant to successfully invade therefore requires

\[
S_{yx}>\frac{N-2}{N-1}(B/2-C)+\frac{1}{N-1}S_{xy}
\]
\\
which in turn implies

\[
B/2-C>S_{xy}
\]
\\
Combining Eqs. 27-29 we can now solve for $v$ and arrive at the following inequality as the condition for a strategy to maintain behavioral diversity in the symmetrical rock-paper-scissors game:

\begin{equation}
p^-_{o} (1-p^-_{-} - p^-_{+} ) >  p^+_{o}(1-p^+_{+} - p^+_{-})
\end{equation}
\\
\bibliographystyle{pnas.bst}

\end{document}